\documentclass{PoS}

\newcommand{\nn}{\nonumber}
\newcommand{\beqn}{\begin{eqnarray}}
\newcommand{\eeqn}{\end{eqnarray}}
\newcommand{\be}{\begin{equation}}
\newcommand{\ee}{\end{equation}}

\newcommand{\ba}{\begin{array}{c}}
\newcommand{\bat}{\begin{array}{cc}}
\newcommand{\ea}{\end{array}}

\newcommand{\bi}{\begin{itemize}}
\newcommand{\ei}{\end{itemize}}

\newcommand{\Frac}[2]{\frac{\displaystyle #1}{\displaystyle #2}}

\newcommand{\cO}{{\cal O}}

\newcommand{\mL}{\mathcal{L}}

\newcommand{\we}{\widetilde{e}}

\newcommand{\bea}{\begin{eqnarray}}
\newcommand{\eea}{\end{eqnarray}}
\newcommand{\bqa}{\begin{eqnarray}}
\newcommand{\eqa}{\end{eqnarray}}
\newcommand{\beq}{\begin{equation}}
\newcommand{\eeq}{\end{equation}}

\newcommand{\bear}{\begin{eqnarray}}
\newcommand{\eear}{\end{eqnarray}}
\newcommand{\ket}{\,\rangle}
\newcommand{\bra}{\langle \,}

\title{$\pi\pi$ scattering lengths at $\cO(p^6)$: resonance estimates}

\ShortTitle{$\pi\pi$ scattering lengths at $\cO(p^6)$}

\author{
Z.H. Guo and \speaker{J.J. Sanz-Cillero}
\ \ \thanks{We would like to thank the organizers of the workshop
for their the attentions and the nice scientific environment.
This talk is based on the study of Ref.~\cite{paper}.
This work has been supported in part by
CICYT-FEDER-FPA2008-01430, SGR2005-00916, the Spanish Consolider-Ingenio 2010
Program CPAN (CSD2007-00042), the Juan de la Cierva program,
the EU Contract No. MRTN-CT-2006-035482
(FLAVIAnet) and National Nature Science Foundation of China under grant number
   10875001, 10575002 and 10721063.
}\\
    Grup de F\'isica Te\`orica and IFAE,
    Universitat Aut\`onoma de Barcelona, 08193 Barcelona, Spain\\
        E-mail:
                \email{zhguo@ifae.es},
                \email{cillero@ifae.es}  }

\abstract{
In a previous paper, some deviations were found in the $\cO(p^6)$
low-energy constants that contribute to the $\pi\pi$--scattering lengths. This
work completes the study of all the relevant couplings ($r_1,\, ... r_6,\, r_{S_2}$).
We also perform a reanalysis of the hadronical inputs used for the estimation
(resonance masses, widths...), checking the impact of the input uncertainties on the
determinations of the chiral couplings and the scattering lengths $a^I_J$.
A good agreement is found with respect to former works,
though our detailed analysis produces a more solid estimate of
these couplings and slightly larger errors.
The effect in the final values of the $a^I_J$ is negligible
after combining them with the other uncertainties, being
the previous scattering length determinations sound and reliable.
Nevertheless, the uncertainties derived here for the $\cO(p^6)$
contributions to the scattering lengths
point out the limitation on further improvements
unless the precision of the $\cO(p^6)$  low-energy couplings is properly increased.
}

\FullConference{International Workshop on Effective Field Theories: from the pion to the upsilon \\
		February 2-6 2009\\
		Valencia, Spain}

\begin{document}
\section{Introduction}

\vspace*{-0.15cm}
This talk presents the culmination~\cite{paper} of a former
work~\cite{matching-op6}, where some of the $\cO(p^6)$
Chiral Perturbation Theory low energy constants
($r_2, \,... \, r_6$)  that describe the $\pi\pi$--scattering were
calculated.
Some shifts were found with respect to former estimates~\cite{bijri},
inducing slight modifications on the
corresponding predictions for the scattering
lengths $a_J^I$ and effective ranges $b_J^I$~\cite{bijri,slgasser}. These constants
provide the partial wave amplitudes for isospin $I$ and angular momentum $J$ near
threshold
$T^I_J(s)\,\, =\,\, k^{2J}\,\, \left(  \, a^I_J\,+\, b^I_J\, k^2\, +...\right)$,
with $k=\sqrt{s/4 -m_\pi^2}$ the pion three-momentum in the dipion
rest-frame~\cite{bijri}.
However, our previous article~\cite{matching-op6} lacked of revised
predictions for the $r_1$  low-energy constant
(LEC), which also enters into the
${   \pi^+(p_1)\pi^-(p_2)\to\pi^0(p_3)\pi^0(p_4)   }$ amplitude at $\cO(P^6)$~\cite{bijri}:
\vspace*{-0.26cm}
\begin{eqnarray}
\label{op6ampop2}
A(s,t,u)|_{r_i}\, = && \,
\frac{m_\pi^4s}{F^6} \left(r_2- 2 r_F\right)
\, + \,  \frac{m_\pi^2s^2}{F^6}r_3
+ \frac{m_\pi^2(t-u)^2}{F^6}r_4
+\frac{s^3}{F^6}r_5
+\frac{s(t-u)^2}{F^6}r_6
+\frac{m_\pi^6}{F^6} \left(r_1+ 2 r_F\right)\, .
\nonumber
\end{eqnarray}
Likewise, the dispersive method considered by Colangelo {\it et al.}~\cite{slgasser}
required  the $\cO(p^6)$ LEC $r_{S_2}$ instead of $r_5$ and $r_6$.
Here we complete the study of these last LECs and perform a full reanalysis of
the different hadronic inputs and their uncertainties.

\vspace*{-0.3cm}
\section{Resonance estimates of $\cO(p^6)$ LECs}

\begin{itemize}

\vspace*{-0.25cm}
\item{} \underline{\bf Set A:}

This is the group of estimates commonly employed in nowadays
calculations~\cite{bijri,bijff}.
The $\chi$PT couplings are assumed to be
determined by the resonance exchanges provided by the phenomenological lagrangian
\vspace*{-0.1cm}
\bear
\mL &=& \Frac{F^2}{4}\bra u_\mu u^\mu +\chi_+\ket
  +\Frac{1}{2}\bra \nabla^\mu S \nabla_\mu S\ket
-\Frac{1}{2}  M_S^{ 2} \bra SS\ket
+c_d \bra S u_\mu u^\mu \ket +c_m \bra S \chi_+\ket
\nn\\
&& \, -\, \Frac{1}{4}\bra \hat{V}_{\mu\nu}\hat{V}^{\mu\nu}\ket
+\Frac{1}{2} M_V^{  2} \bra \hat{V}_\mu\hat{V}^\mu\ket \,\,
-\Frac{i g_V}{2\sqrt{2}}\bra \hat{V}_{\mu\nu}[u^\mu,u^\nu]\ket
+f_\chi \bra \hat{V}_\mu [u^\mu,\chi_-]\ket
 \, ,
\label{lagr}
\eear
where $\bra ...\ket$ stands for trace in flavour space, $S$ and $\hat{V}^\mu$
account  respectively for the scalar and vector multiplets. The tensor
$u^\mu$ contains the chiral pseudo-Goldstone and $\chi_\pm$
is, in addition, proportional to the light quark masses. Their precise definitions
can be found in Refs.~\cite{bijri,bijff,ecker89}. From the comparison of the
$\rho\to\pi\pi$ and $K^* \to K\pi$ decays and other processes,
Ref.~\cite{bijri} obtained the set of parameters
\vspace*{-0.25cm}
\bear
M_V=770\, \mbox{MeV}\, , \qquad
g_V=0.09\, , \qquad f_\chi=-0.03  \, ,
\nn\\
M_S=983\, \mbox{MeV}\, , \qquad c_m=42\, \mbox{MeV}\, ,\qquad
c_d=32\, \mbox{MeV}\, .
\label{setAinputs}
\eear
Taking this inputs and the phenomenological lagrangian~(\ref{lagr}),
 Ref.~\cite{bijri} provided
\vspace*{-0.25cm}
\bear
r_1^A=-0.6 \times 10^{-4}\, , \qquad  &
r_2^A=1.3 \times 10^{-4}\, , \qquad
&r_3^A=-1.7 \times 10^{-4}\, ,
\label{eq.setA}
\\
r_4^A=-1.0 \times 10^{-4}\, , \qquad  &
r_5^A=1.1 \times 10^{-4}\, , \qquad
&r_6^A=0.3 \times 10^{-4}\, , \qquad
r_{S_2}^A\, =\,  -0.3 \times 10^{-4}\, .
\nonumber
\eear
\vspace*{-0.35cm}

\vspace*{-0.6cm}
\item{} \underline{\bf Set B:}

However,  some scalar meson contributions were found to be missing
in previous estimates of the $\cO(p^6)$ LECs~\cite{bijri,slgasser,bijff}.
The couplings $r_2,\, ... \, r_6$ were fully calculated at large $N_C$~\cite{matching-op6},
being expressed in terms of the ratios
\begin{eqnarray}
\frac{\Gamma_R}{M_R^3} = \frac{\overline{\Gamma}_R}{\overline{M}_R^{\,\, 3}} \left[1+
\alpha_R\frac{m_\pi^2}{\overline{M}_R^{\,\, 2}}
+\gamma_R\frac{m_\pi^4}{\overline{M}_R^{\,\, 4}}+\cO(m_\pi^6)\right]\, ,
\quad \,\,\,\,
\label{beta}
\frac{\Gamma_R}{M_R^5}  = \frac{\overline{\Gamma}_R}{\overline{M}_R^{\,\, 5}} \left[1+
\beta_R\frac{m_\pi^2}{\overline{M}_R^{\,\, 2}}+\cO(m_\pi^4)\right] ,
\label{KSRFdefi}
\end{eqnarray}
where $\overline{M}_R$ and $\overline{\Gamma}_R $ stand for the chiral
limit of $M_R$ and $\Gamma_R$, respectively. The constants
$\alpha_R$, $\beta_R$, $\gamma_R$ are quark mass independent and
rule the $m_\pi$  corrections in the ratios.
The resonance masses and widths were computed at large $N_C$
by means of the resonance lagrangian~(\ref{lagr}).
Using exactly the same inputs (\ref{setAinputs}) of set A,
we found that $r_5$ and $r_6$ remained unchanged, $r_3$ and $r_4$ varied slightly
($r_3^B=0.9  \times 10^{-4} , \,  r_4^B=-1.9 \times 10^{-4}$)
and $r_2$ suffered a big variation
($r_2^B=18  \times 10^{-4}$)~\cite{paper,matching-op6}.

\vspace*{-0.25cm}
\item{} \underline{\bf Set C:}

\vspace*{-0.15cm}
As relevant variations were found in some of the $r_i$, in addition to performing
the full large--$N_C$ estimate (without dropping any possible resonance contribution),
a detailed analysis of the experimental inputs and their uncertainties
was also in order.
It was  found that although the vector sector is quite under control,
our knowledge on the scalar resonance properties is rather poor. This work
is devoted to this analysis.

\end{itemize}

\section{Phenomenology of the resonance parameters}
\label{pheno}

\vspace*{-0.2cm}
\subsection{Mass splitting up to $\cO(m_P^2)$}

In the large--$N_C$ limit, the mass splitting
of the resonance multiplets can be described at leading order by one single
operator $e_m^R$~\cite{mass-split}
\bqa
\label{eq.eRm}
 -\frac{\overline{M}_R^{\,\,2}}{2}\bra R R\ket \, + \, e_m^R \bra R R \chi_+\ket \,,
\eqa
which leads at large $N_C$ to the mass eigenstates
\bqa
M^2_{I=1} &=& \overline{M}_R^{\,\,2} - 4 e_m^R m_\pi^2\,\,\,
+\,\,\, \cO(m_P^4) \,\,=\,\,
M^{(\bar{u}u+\bar{d}d)}_{I=0}   \,,
\nonumber\\
M^2_{I=\frac{1}{2}} &=& \overline{M}_R^{\,\,2} - 4 e_m^R m_K^2\,\,\,
+\,\,\,\cO(m_P^4)  \,,
\nonumber\\
M^{(\bar{s}s)\,\,\, 2}_{I=0} &=& \overline{M}_R^{\,\,2}
- 4 e_m^R \,\, (2 m_K^2-m_\pi^2)\,\,\,
+\,\,\,\cO(m_P^4)  \,.
\label{masssplit}
\eqa

The combined study of the $\rho(770)$, $K^*(892)$ and $\phi(1020)$ masses
leads to the values~\cite{paper}
\be
\label{eV2}
\overline{M}_V \, \,= \,\,  764.3 \pm 1.1  \, \mbox{MeV}\, ,
\qquad
e_m^V\, \, = \,\,  -0.228\pm 0.015 \,.
\ee

In the case \  of the scalars, \ the lightest  \ $I=1$ resonance is \  identified with the $a_0(980)$:
${   M_{I=1}=  984.7\pm 1.2 }$~MeV~\cite{pdg}.
In order to avoid the problem
of the mixing of iso-singlet scalars, the analysis is performed with
the $I=1/2$ state.
The broad  $\kappa(800)$ seems to be a possible candidate although the first
clear $I=1/2$ scalar resonance signal is provided by the $K^*_0(1430)$~\cite{pdg}.
Hence, we take the conservative estimate
$M_{I=1/2}= 1050\pm 400$~MeV, which ranges from the $\kappa$
up to the $K^*_0(1430)$ mass.  This leads then to the values
\bqa
\label{eS}
\overline{M}_S \,=\, 980 \pm 40 \, {\rm{MeV}}\,,\qquad e_m^S\,=\, -0.1 \pm 0.9 \,.
\eqa

\subsection{The splitting of the vector resonance  decay width up to $\cO(m_P^2)$}
\label{avsection}

The vector  decay width into two light pseudo-scalars,
$V \to \phi_1 \phi_2$,  shows the   general structure
\bqa
\label{avdef}
\Gamma_{V\to\phi_1\phi_2} &=& C_{V 1 2}
\,\, \times\,\,
\Frac{ M_V^{\,\, n} \, \rho_{\rm V 12}^3 }{48\, \pi \,  F_{1}^2 \, F_{2}^2}
\,\,\, \lambda_{V\pi\pi}^2\,\,\, \left[1 + \epsilon_V \frac{m_1^2+m_2^2}{2 \overline{M}_V^2}
\,\,\,+\,\,\, \cO(m_P^4)\right]^2\,,
\eqa
with the phase-space factor
$\rho_{\rm V 12}= M_V^{-2} \sqrt{(M_V^2-(m_1+m_2)^2 )
\,(M_V^2-(m_1-m_2)^2)}$.
The $F_i$ are the physical decay constants  for the $\phi_i$
pseudo-Goldstones ($F_\pi\simeq 92.4$~MeV
and $F_K\simeq 113$~MeV)  and they appear due to the
large--$N_C$ wave function renormalization of the light
pseudo-scalars~\cite{matching-op6,juanjotadp}. $M_V$ and $m_i$ correspond, respectively,
to the physical vector and pseudo-scalar mass.
Depending on the channel, one has the Clebsch-Gordan $C_{\rho\pi\pi}=1$,
$C_{K^* K\pi}=3/4$ and $C_{\phi K\overline{K}}=1$.
The $V\phi_1 \phi_2$ coupling and the mass scaling $M_V^n$ depend on the
considered lagrangian realization, either
Proca fourvector (n=5) or Antisymmetric tensor formalism (n=3)~\cite{ecker89,bijff,spin1fields}.
The combination of the experimental $K^*$ and $\rho$ widths yields~\cite{paper}:
\bear
\lambda_{V\pi\pi}=g_V=
0.0846 \pm 0.0008\, ,& \qquad \quad
 \epsilon_V^{\rm  }=0.01 \pm 0.09\, \,,\qquad
&\mbox{( Proca~\cite{bijff,spin1fields} ) ,}
\\
\lambda_{V\pi\pi}=G_V = 63.9 \pm 0.6 \, \mbox{MeV} \, ,& \qquad \quad
\epsilon_V^{\rm  } =0.82\pm 0.10 \, , \qquad
&\mbox{( Antisym.~\cite{ecker89,spin1fields} ) .}
\nonumber
\eear

\subsection{The decay width for the scalar resonance}

In the case of the scalar mesons the current knowledge
nowadays is still very poor.
We had then to rely on the phenomenological lagrangian~(\ref{lagr}) for the
description of the $a_0(980)\to \pi\eta$ width~\cite{paper,ecker89}, and on
the theoretical scalar form-factor constraint
$4 c_d c_m=F^2$~\cite{cdcm}:
\bqa
c_d\, =\,   26 \pm 7  \, {\rm{MeV}}\,,\qquad c_m\, =\, 80 \pm 21 \,{\rm{MeV}}\,,
\eqa
where their large errors stems essentially from the wide range
we considered for the $a_0(980)$ partial width,
$\Gamma_{a_0\to\pi\eta}=75\pm 25$~MeV~\cite{paper}.

\subsection{Chiral corrections to $F_\pi$}
\label{sec.Fpi}

At large--$N_C$, the   wave-function renormalization  of the $\pi$ field
is related to the   decay constant
in the way $F_\pi=F\, Z_\pi^{-1/2}$ ~\cite{juanjotadp,bernardtadp}.
The $m_\pi^2$ corrections to $F_\pi$ can be parametrized  in the form
\be
\label{Fpi-F}
F_\pi\,\,\, =\,\,\, F\,\, \left[ \, 1
\,\, +\,\, \delta F_{(2)}\, \Frac{m_\pi^2}{\overline{M}_S^{\,\, 2}}
\,\, +\,\, \delta F_{(4)}\, \Frac{m_\pi^4}{\overline{M}_S^{\,\, 4}}
\,\, +\,\, \cO(m_\pi^6)\, \right]\, .
\ee
The scalar lagrangian (\ref{lagr}),
the mass splitting~(\ref{masssplit}) and the former inputs produce the predictions
\be
\delta F_{(2)}=\Frac{4 c_d c_m}{F^2}\,=\, 1\, ,
\qquad \qquad
\delta F_{(4)}= \Frac{8 c_d c_m}{F^2}\left(
\Frac{3 c_d c_m}{F^2}-\Frac{4 c_m^2}{F^2}\right)\,
+  \Frac{16 c_d c_m e_m^S}{F^2}\,=\, -5\pm 5 \, .
\label{eq.Fpi-ecker}
\ee

\subsection{Next-to-next-to-leading order
chiral corrections to $M_V$, $\Gamma_V$ and $\Gamma_S$}

The next-to-next-to-leading order  corrections
(NNLO)   to the
vector mass are also needed in order to extract the LEC $r_2$~\cite{paper}.
At large $N_C$, the quark mass corrections are given at NNLO  by
$ {      M_{I=1}^2 = \overline{M}_R^2 -  4 e^R_m m_{\pi}^2 -  4 \we^{R}_{m}
m_{\pi}^4 /\overline{M}_R^{\,\, 2}      }$.
Demanding   that the NNLO terms never overcome the NLO corrections
in the vector multiplet  sets the range
$ \left| \we^{V}_{m} \right|
\,\, \leq \,\,
 \Frac{ \overline{M}_V^{\,\, 2}}{2 m_K^2-m_\pi^2}\,\,
 |e^V_m|  \,\,\,\,
 \simeq \,\,\,\, 0.3$~\cite{paper}.

The determination of $r_2$ also requires the   NNLO chiral corrections
$\widetilde{\epsilon}_R$ to the resonance widths
\vspace*{-0.15cm}
\bear
\Gamma_{\rho\to\pi\pi} &=&
\Frac{ M_\rho^{\,\, n} \, \rho_{\rho\pi\pi}^3 }{48\, \pi \,  F_{\pi}^4}
\,\,\, \lambda_{V\pi\pi}^2\,\,\, \left[1 + \epsilon_V \Frac{m_\pi^2}{\overline{M}_V^{\,\, 2}}
\,
+ \,  \widetilde{\epsilon}_V\Frac{m_\pi^4}{\overline{M}_V^{\,\, 4}} \,\,\,+\,\,\, \cO(m_\pi^6)\right]^2\,,
\nn\\
\Gamma_{\sigma\to \pi\pi} \,\,&=&\,\,
\Frac{3 M_\sigma^{3} \rho_{\sigma\pi\pi}}{16\pi F_\pi^4} \,\,\,
c_d^2 \, \left[1  +  \epsilon_S \Frac{m_\pi^2}{\overline{M}_S^{\,\,2}}
+  \widetilde{\epsilon}_S\Frac{m_\pi^4}{\overline{M}_S^{\,\, 4}}
+ \cO(m_\pi^6)\right]^2\, .
\eear
The phenomenological lagrangian (\ref{lagr})~\cite{ecker89,bijff} yields the
predictions
\bqa
\widetilde{\epsilon}_V&=&\epsilon_V \,\left[ \Frac{8  c_m(c_d-c_m)}{F^2}
\,\Frac{\overline{M}^2_V }{\overline{M}^{\,\, 2}_S}\, +
4 e_m^V \right]
\,\, =\,\, \left\{\begin{array}{ll} -0.03\pm 0.18\, , \qquad&\mbox{(Proca)} \\
-1.6\pm 0.9\, ,\qquad&\mbox{(Antisym.)}  \end{array}\right.
\nonumber \\
\widetilde{\epsilon}_S&=&\Frac{16 c_m^2(c_d-c_m)}{c_d F^2}
+ \Frac{8 (c_m-c_d) e_m^S }{c_d }\,\,=\,\, -7\pm 12 \,.
\eqa

\section{Low-energy constant determination at $\cO(p^6)$}

Based on the partial-wave dispersion relations developed in
Refs.~\cite{matching-op6,matching}, it is possible to extract
the large--$N_C$ values of $r_2\, ,...\, r_6$ from the
$I=1$ vector  and $I=0$ scalar  $(\bar{u} u +\bar{d} d)$  width and mass ratios
$\Gamma_R/M_R^3$ and $\Gamma_R/M_R^5$:
the couplings $r_5$ and $r_6$ are determined by the $\Gamma_R/M_R^5$ ratio
in the chiral limit; $r_3$ and $r_4$ also require its first $m_\pi^2$ correction
$\beta_R$; those and the NNLO $m_\pi^2$ contribution to $\Gamma_R/M_R^3$
are needed in order to obtain $r_2$.
All these LECs have been found to be dominated by the vector resonance exchanges.
The $\cO(p^6)$ couplings $r_1$~\cite{bijff} and $r_{S_2}$~\cite{bijff} could not be computed
through the partial-wave dispersion relations in~\cite{matching-op6,matching}.
They were calculated directly from the phenomenological lagrangian~(\ref{lagr}):
\bear
r_1^{\rm Proca}&=&  -\Frac{16 c_d c_m ( 8 c_d^2-17 c_d c_m + 12 c_m^2)}{\overline{M}_S^{\,\,4}}
+ \Frac{32(c_d-c_m)^2 F^2}{\overline{M}_S^{\,\,4}}\, e_m^S
\nn\\
&&\qquad
-\Frac{16 g_V^2 F^2}{\overline{M}_V^{\,\,2} }
\left[1 + \epsilon_V +\Frac{1}{4}  \epsilon_V ^2
- \Frac{ 8 c_d c_m}{F^2} \Frac{\overline{M}_V^{\,\,2}}{\overline{M}_S^{\,\,2}} \right] \, ,
\label{eq.r1Proca}
\\
r_{S2} &=&  \Frac{ 8 c_m (c_m-c_d) F^2}{\overline{M}_S^{\,\,4}}
\,-\, \Frac{32 c_d^2 c_m^2}{\overline{M}_S^{\,\,4}}
+ \frac{16c_d c_m F^2}{\overline{M}_S^{\,\,4}}e_m^S\,.
\eqa
The expression for $r_1$ in the Antisymmetric tensor formalism is similar to
(\ref{eq.r1Proca}) but with the second line replaced by
$ -\frac{16 G_V^2F^2}{\overline{M}_V^{\,\,4}} \left[ 1+ \epsilon_V
- \frac{ 8 c_d c_m}{F^2} \frac{\overline{M}_V^{\,\,2}}{\overline{M}_S^{\,\,2}}
+ 2 e^V_m  \right]$.
All this leads to the values of the low-energy constants shown in Table~\ref{tab.ri}.
The first error derives from the phenomenological inputs and the second one
stems from the uncertainty on the saturation scale $\mu_s$ where
$r_i^r(\mu_s)=r_i^{N_C\to\infty}$~\cite{paper}.

\begin{table}[!h]
\centering
\begin{tabular}{|c|c|c|c|c|}
  \hline
 & ND est.~\cite{op6-reno} & set A~\cite{bijri,bijff}   & set C  (Proca)
 & set C (Antisym.)
\\
    \hline
  $10^4 \cdot r_1^r$ & $\pm 80$ &$-0.6$  &
  $-14\pm 17\pm 3$  &
  $-20\pm 17 \pm 3$
  \\
  $ 10^4 \cdot r_2^r$& $\pm 40$ & $1.3$ &
  $22\pm 16\pm 4  $   &  $ 7\pm 10\pm 4 $
  \\
  $10^4 \cdot r_3^r$ & $\pm 20$ & $-1.7$ &
  $-3\pm 1 \pm 3$    & $ -4\pm 1  \pm 3$
  \\
  $10^4 \cdot r_4^r$ & $\pm 3$ &$-1.0$ &
  $-0.22\pm 0.13\pm 0.05$     & $ 0.13\pm 0.13\pm 0.05$
  \\
  $10^4 \cdot r_5^r$ & $\pm 6$ & $1.1$  &
  $0.9\pm 0.1\pm 0.5 $     & $0.9\pm 0.1 \pm 0.5$
  \\
  $10^4 \cdot r_6^r$ & $\pm 2$ & $0.3$  &
  $0.25\pm 0.01\pm 0.05$    & $ 0.25\pm 0.01\pm 0.05$
  \\
  $10^4 \cdot r_{S_2}^r$ & $\pm 1$ & $-0.3 $  &
  $1\pm 4 \pm 1 $    & $1\pm 4 \pm 1 $
  \\
  \hline
\end{tabular}
\caption{{\small
Different predictions for the $\cO(p^6)$ LECs  $r_i^r(\mu)$ for $\mu=770$~MeV:
The first column presents the order of magnitude estimate based on
naive dimensional analysis~\cite{op6-reno};
In the set A column  we show former estimates from Refs.~\cite{bijri,bijff};
in the last two columns,
one can find the values for the present reanalysis.}}
\label{tab.ri}
\end{table}

\vspace*{-0.25cm}
\section{Scattering lengths }

The $\chi$PT  expression for the scattering lengths contains at $\cO(p^6)$
a series of logarithmic terms  together
with analytical $\cO(p^6)$ contributions~\cite{bijri}.
Although the first ones are the most complicate contributions to compute,
their value is nevertheless rather sound and under control. On the other
hand, the local terms are determined by the $\cO(p^6)$ LECs $r_i$ and, although they
can be easily computed, these couplings are badly known
and their estimation is pretty cumbersome.

In Ref.~\cite{slgasser}, Colangelo {\it  et al.} combined the NNLO chiral
perturbation theory computation of the scattering lengths~\cite{bijri}
with a  phenomenological dispersive representation.  This allowed them to produce
one of the most precise determinations of the scattering lengths.
They were expressed in terms of some dispersive integrals,
the pion quadratic scalar radius $\langle r^2\rangle_S^\pi$,
the $\cO(p^4)$ coupling $\ell_3$ and a set of $\cO(p^6)$ LECs
($r_1,\ r_2,\, r_3,\, r_4$, $r_{S_2}$).
Following the work of Ref.\cite{slgasser},
we extracted the part of their scattering lengths  that depended
on the inputs  $r_i^r(\mu)$~\cite{paper,slgasser}:
\vspace*{-0.15cm}
\bqa
a^0_0|_{r_i} =& \frac{7m_\pi^2}{32\pi F_\pi^2}  C_0|_{r_i}\, &=\,
\Frac{m_\pi^6}{32 \pi F_\pi^6}
\left[ 5 r_1^r + 12 r_2^r + 28 r_3^r - 28 r_4^r -14 r_{S_2}\right] \, ,\quad
\nonumber \\
a^2_0|_{r_i}  =& -\frac{m_\pi^2}{16\pi F_\pi^2}  C_2|_{r_i}
\,&=\, \Frac{m_\pi^6}{16\pi F_\pi^6}
\left[ r_1^r - 4r_3 + 4 r_4 + 2 r_{S_2}\right]
\,.
\label{scalgasser}
\eqa

The largest contributions to the $a_0^0$ and $a^2_0$ errors are found to be
produced in similar terms by
$r_1$, $r_2$, $r_3$ and $r_{S_2}$, being   the impact of $r_4$  negligible.

\begin{table}[!h]
\centering
\begin{tabular}{|c|c|c|c|c|}
  \hline
 & Total: Ref.~\cite{slgasser}  &  $a^I_J|_{r_i}$~\cite{slgasser}
 & $a^I_J|_{r_i}$~Set C    (Proca) & $a^I_J|_{r_i}$~Set C   (Antisym.)
   \\
  & $\qquad ( \times 10^{-3}) \qquad $ & $\qquad ( \times 10^{-3}) \qquad $
  &  $ ( \times 10^{-3})$  & $(\times 10^{-3})$
   \\
    \hline
     \hline
  $a^0_0$ &  $220\pm 5$ & $0.0\pm 1.0 $  &  $1.0 \pm 1.5 \pm 1.0 $
  & $-1.6 \pm 1.5 \pm 1.0$
  \\
  $10 \, a^2_0$ & $-444\pm 10$  & $0.4\pm 2.0 $ &   $0 \pm 4\pm 2$
  & $0\pm 4\pm 2$
  \\
  \hline
\end{tabular}
\caption{{\small   The first and second columns  show, respectively,
the total scattering lengths
and the $r_i$ contribution to them in the dispersive method from
Colangelo {\it et al.}~\cite{slgasser},
where the authors used the $r_i$ in Eq.~(2.3),
 $F_\pi=92.4$~MeV and $m_\pi=139.57$~MeV.
The last two columns  show the reanalyzed quantities $a^I_J|_{r_i}$ (set C)  for
the Proca   and antisymmetric tensor formalisms
for the usual scale $\mu=770$~MeV. There,
the first error derives from the inputs and the second one from the
saturation scale uncertainty.
}}
\label{tab.CGL}
\end{table}

\vspace*{-0.35cm}
\section{Summary and conclusion}

The combination of  the Proca and antisymmetric results
yields for our prediction of the LECs the final numbers (for $\mu=770$~MeV),
\bear
r_1^r=(-17\pm 20) \times 10^{-4}\, , \qquad  &
r_2^r=(17\pm 21)\, \times 10^{-4}\, , \quad
&r_3^r=(-4\pm 4)\,  \times 10^{-4}\, ,
\nn \\
r_4^r=(0.0\pm 0.3)\,  \times 10^{-4}\, , \qquad  &
r_5^r=(0.9 \pm 0.5 )\,  \times 10^{-4}\, , \quad
&r_6^r=(0.25 \pm 0.05)\, \times 10^{-4}\, ,
\nn\\
r_{S_2}^r=(1\pm 4)\,  \times 10^{-4}\, . \qquad  &
\eear
The $r_i^r$ contributions to
the scattering lengths with the Colangelo {\it et al.}'s  method~\cite{slgasser}
can be summarized in the predictions
\vspace*{-0.25cm}
\bqa
10^3 \, a^0_0|_{r_i}\, =\, 0 \pm 3 \, ,
\qquad\qquad
10^4  a^2_0|_{r_i} \, =\, 0 \pm 5 \, .
\label{eq.aIJ_ri}
\eqa
Following \ the  \ analysis  \ of \  global \  uncertainties \
of  \ Ref.~\cite{slgasser}  \ leads  \ to the updated values  \
$ {   a^0_0 =  0.220 \pm 0.005    }$ and
$10\,  a^2_0  =  -0.444\pm 0.011 $.
These values leave essentially unchanged the previous determinations
$a_0^0=0.220\pm 0.005$ and $10\,  a^2_0=-0.444\pm 0.010$~\cite{slgasser}.

This calculation shows that   the determinations of the scattering lengths
through the dispersive method and  $r_i$ resonance saturation
estimates are rather solid~\cite{slgasser}.
Our detailed analysis shows that the  error stemming from the $r_i$ does not
modify the final numbers quoted in Ref.~\cite{slgasser}.
Nonetheless, unless the the precision in the $\cO(p^6)$ low-energy constants is
conveniently increased, it will be difficult to carry on further relevant
improvements in the scattering length determinations.


\vspace*{-0.25cm}
\begin{thebibliography}{99}

\bibitem{paper}
    Z.-H. Guo and J.J. Sanz-Cillero,
    {\it $\pi\pi$--scattering lengths at $O(p^6)$ revisited},
    [arXiv:0903.0782 [hep-ph]]  (Phys.~Rev.~D in press).

\bibitem{matching-op6}
    Z.~H.~Guo, J.~J.~Sanz Cillero and H.~Q.~Zheng,
    {\it $\cO(p^6)$ extension of the large--$N_C$  partial-wave dispersion relations}\,,
    Phys. Lett. {\bf B 661} (2008) 342.




\bibitem{bijri}
    J.~Bijnens, G.~Colangelo, G.~Ecker, J.~Gasser and M.~E.~Sainio,
    {\it  Pion pion scattering at low-energy},
    Nucl.Phys.{\bf B 508} (1997) 263, Erratum-ibid.{\bf B 517}(1998) 639.


\bibitem{slgasser}
    G.~Colangelo, J.~Gasser and H.~Leutwyler,
    {\it $\pi\pi$ scattering},
    Nucl. Phys.{\bf B 603} (2001) 125.

\bibitem{bijff}
    J.~Bijnens, G.~Colangelo and P.~Talavera,
    {\it The Vector and scalar form-factors of the pion to two loops},
    JHEP {\bf 05} (1998) 014.


\bibitem{ecker89}
    G. Ecker {\it et al.},
    {\it The Role of Resonances in Chiral Perturbation Theory},
    Nucl. Phys. {\bf B 321} (1989)311.


\bibitem{mass-split}
    V. Cirigliano, G. Ecker, H. Neufeld and A. Pich,
    {\it  Meson resonances, large--$N_C$ and chiral symmetry},
    JHEP {\bf 0306} (2003) 012.


\bibitem{pdg}
    C. Amsler {\it et al.} (Particle Data Group),
    {\it Review of Particle Properties},
    Phys. Lett. {\bf B 667} (2008) 1.


\bibitem{juanjotadp}
    J.J. Sanz-Cillero,
    {\it Pion and kaon decay constants: Lattice versus resonance chiral theory},
    Phys. Rev. {\bf D 70} (2004) 094033.



\bibitem{bernardtadp}
    V. Bernard, N. Kaiser and Ulf G. Meissner,
    {\it Chiral perturbation theory in the presence of
    resonances: Application to $\pi\pi$ and $\pi K$ scattering},
    Nucl. Phys. {\bf B 364} (1991) 283.


\bibitem{spin1fields}
  G.~Ecker, J.~Gasser, H.~Leutwyler, A.~Pich and E.~de Rafael,
  {\it Chiral Lagrangians for massive spin 1 fields},
  Phys.\ Lett.\   {\bf B 223}, 425 (1989).



\bibitem{matching}
    Z.~H.~Guo, J.~J.~Sanz Cillero and H.~Q.~Zheng,
    {\it Partial-waves and large--$N_C$ resonance sum rules},
    JHEP {\bf 0706} (2007) 030.

\bibitem{cdcm}
    M. Jamin, J.A. Oller and A. Pich,
    {\it Strangeness changing scalar form-factors},
    Nucl.~Phys.~{\bf B 622} (2002) 279.


\bibitem{op6-reno}
    J. Bijnens, G. Colangelo and G. Ecker,
    {\it Renormalization of chiral perturbation theory to order $ p^6 $},
    Annals Phys. {\bf 280} (2000) 100-139.


\end{thebibliography}
\end{document}